Comment on "[N(CH$_3$)$_3$H]$_2$ZnCl$_4$: Ferroelectric properties and characterization of phase transitions by Raman spectroscopy"

Zbigniew Tylczyński

Faculty of Physics, Adam Mickiewicz University, Umultowska 85, 61-614 Poznań, Poland

E-mail address: zbigtyl@amu.edu.pl

Keywords: ferroelecticity, dielectric constant, ac conductivity

The authors of paper commented on measured dielectric constant (on pellet sample only) and Raman spectra of trimethylammonium tetrachlorozincate crystal – [N(CH$_3$)$_3$H]$_2$ZnCl$_4$ – in the temperature range 220–385 K.[1] The symmetry and structure of this substance was resolved by Williams and Brown in 1992 for monocrystals.[2] The authors claim that their crystal shows a rich series of phase transitions similarly to [N(CH$_3$)$_4$]$_2$ZnCl$_4$ crystal. But no such phase transitions were found in the trimethylammonium compound since hydrogen bonding apparently locks-in the commensurate phase.[2]

The authors wrote that "a predominant dielectric peak at T$_2$ = 282 K suggests the presence of ferroelectric-paraelectric phase transition". If that were true, the spontaneous polarization at the low-temperature phase should arise as a macroscopic order parameter. Moreover, the polarization vector will have to be reversed by external electric field giving a characteristic hysteresis loop. This is the classical definition of ferroelectricity.[3] In the commented paper no such loop had be observed. In the low-temperature phase the ferroelectric domain walls should exist giving dielectric relaxation in a low frequency electric field. One cannot see this relaxation in figure 5. The Curie-Weiss law was claimed to be satisfied on the basis of only 3 points of temperature dependence of 1/ε' – figure 7. In this figure T$_C$=T$_0$, from which the authors conclude that the phase transition is of the second order. This conclusion is contrary to the DSC data in figure 3 where the phase transition at 282 K has a strong first order character. Moreover, what is the point group of symmetry in the low-temperature phase?

At many different phase transitions the dielectric anomaly can exist due to, for example, change in conductivity, especially in the crystals having many hydrogen bonds. Moreover, in the whole measured temperature range the dielectric loss (ε'') is 100 times higher than the real part of dielectric constant (ε'). This situation takes place at T$_4$ too. From figure 5 and previous authors' paper one can calculate *ac* conductivity as 2.3·10$^{-2}$ (Ωm)$^{-1}$ for 11 kHz.[4] This value is characteristic of high conducting materials, not of dielectric crystals. The spontaneous polarization, *i.e.* the stable electric charges on the opposite surfaces of crystal, does not exist in such high conducting crystal.